\documentclass[12pt]{article}
\newcommand{\stln}{\setlength{\unitlength}{2.3 ex}}

\newcommand{\sfr}{\framebox(1,1){\begin{picture}(1,1)
  \put(0,0){}\end{picture}}}
\newcommand{\lmarcshapirowbox}
{\stln \lower2.6ex\hbox{
\begin{picture}(6.9,2.6)
\multiput(.3,1.3)(1,0){2}{\sfr}
\put(2.3,1.3){\framebox(3,1){$\cdots$}}
\put(5.3,1.3){\sfr}
\multiput(.3,.3)(1,0){2}{\sfr}
\put(2.3,.3){\framebox(3,1){$\cdots$}}
\put(5.3,.3){\sfr}
\put(.3,0){$\underbrace{~~~~~~~~~~~~~~~~~~}_{j}$}
\end{picture}}}
\font\mybb=msbm10 at 12pt
\let\a=\alpha\let\b=\beta
\let\e=\epsilon

\newcommand{\nn}{\nonumber}

\newcommand{\be}{\begin{equation}}
\newcommand{\ee}{\end{equation}}
\newcommand{\bea}{\begin{eqnarray}}
\newcommand{\eea}{\end{eqnarray}}
\newcommand{\eps}{\epsilon}

\renewcommand{\paragraph}[1]{
\vspace{.8mm}\par\noindent {\sl #1}\\
\vspace{0.2mm} }

\newcommand{\eqn}[1]{(\ref{#1})}

\newcommand{\ba}{\left(\begin{array}}
\newcommand{\ea}{\end{array}\right)}

\def\da{{\dot\alpha}}
\def\db{{\dot\beta}}

\newsavebox{\uuunit}
\sbox{\uuunit}
{\setlength{\unitlength}{0.825em}
 \begin{picture}(0.6,0.7)
\thinlines
\put(0,0){\line(1,0){0.5}}
\put(0.15,0){\line(0,1){0.7}}
 \put(0.35,0){\line(0,1){0.8}}
\multiput(0.3,0.8)(-0.04,-0.02){12}{\rule{0.5pt}{0.5pt}}
\end {picture}}
\newcommand {\unity}{\mathord{\!\usebox{\uuunit}}}
\begin{document}
\begin{titlepage}
\begin{flushright}
SU-ITP-99/30\\
KUL-TF-99/25\\
{\tt hep-th/9906195}\\
June 25, 1999\\
\vskip 1 cm
{\it Dedicated to the memory of Efim Samoilovich Fradkin}
\end{flushright}
\vskip 1.3cm
\begin{center}
{\large {\bf BRST Quantization of a  Particle in AdS$_5$}}
\vskip 1.5cm
{\bf  Piet Claus$^{\dagger a}$, Renata Kallosh$^{* b}$ and J.~Rahmfeld$^{*
c}$}\\
\vskip.3cm
{\small
$^\dagger$
Instituut voor theoretische fysica, \\
Katholieke Universiteit Leuven, B-3001 Leuven, Belgium\\
\par
\
\par
$^*$ Physics Department, \\
Stanford University, Stanford, CA 94305-4060, USA
}
\vskip 0.6cm
\begin{abstract}

We perform the quantization of a massive particle propagating on $AdS_5$.
We use the twistor formulation in which the action can be brought into a
quadratic form. We construct the BRST operator which commutes with $AdS_5$
isometries forming $SU(2,2)$. The condition of a consistent BRST quantization
requires that the AdS energy $E$ is quantized in units
of the $AdS_5$ radius $R$, $E=\frac{1}{2R}(N_a +N_b+4 )$,
with $N_a,  N_b$ being some non-negative integers. We also argue that the mass
operator will be identified with  the moduli of the $U(1)$ central extension
$Z$ of the $SU(2,2|4)$ algebra in the supersymmetric case.
The spectrum of physical states with vanishing ghost number
contains a particular subset of `massless' $SU(2,2)$ multiplets
(including  the bosonic part of the `novel short' supermultiplets).
We hope that our results will help to quantize also the
string on $AdS_5$.
\end{abstract}

\end{center}

\vfill
\footnoterule \noindent
{\footnotesize $\phantom{a}^a$ e-mail: piet.claus@fys.kuleuven.ac.be.}\\
{\footnotesize $\phantom{b}^b$ e-mail: kallosh@physics.stanford.edu.}\\
{\footnotesize $\phantom{c}^c$ e-mail: rahmfeld@leland.stanford.edu.}

\end{titlepage}

The quantization of string theory on $AdS_5\times S^5$ is a
complicated problem.  One of the main difficulties is due to the
non-linearities,  present even in the bosonic part. The action  is not
quadratic
and hence difficult to quantize.  It has been first suggested \cite{CGKRZ} to
use the supertwistors forming the fundamental representation of $SU(2,2|4)$
as the fundamental variables of the theory, as they should linearize
the symmetries of the string action.
A toy model  world-sheet action using supertwistor variables
was suggested and its  quantization  resulted in the spectrum
of states forming unitary irreducible representations of $SU(2,2|4)$.
The connection of this supertwistor action with linearly realized
$SU(2,2|4)$ to the gauge-fixed string action with non-linearly realized
$SU(2,2|4)$ was not clear however. Progress was made in \cite{CRZ}
where a twistor parametrization of $AdS_5$ was found. The results
were applied to a massive particle propagating on $AdS_5$, and it was
shown that the resulting action is a simple quantum mechanical system
with an $U(1)\times SU(2)$ gauge symmetry and a linear
$SU(2,2)$ global symmetry. Of course, the goal is  eventually to use
the results for a quantization of strings on $AdS_5\times S^5$, thus
solving a long-standing problem.
In this paper we consider again the massive particle on $AdS_5$.
Already in the particle case the problems analogous to those in string theory
exist. We consider the action of \cite{CRZ} and argue by counting of the number
of
degrees of freedom that quantum-mechanically the twistor version is equivalent
to the space-time  version of the particle theory propagating in $AdS_5$
space.
We then find a gauge leading to  the action in a quadratic form and
perform its BRST quantization, thus giving hope and further support for the
program of the quantization of the string.

The action of a massive particle propagating in AdS$_5$ space is proportional
to the invariant length of the world line
\begin{equation}
S= -m \int ds
\label{particlesimple}
\end{equation}
where
\begin{equation}
-ds^2 = d\tilde x^{\tilde m} \tilde g_{\tilde m\tilde n} d\tilde x^{\tilde n}
= \rho^2 dx^2 + R^2 \,\left(\frac{d\rho}\rho\right)^2\, .
\label{adsmetric}
\end{equation}
This action can  also be presented in an equivalent form
\begin{equation}
S = \int d\tau \left[\tilde P_{\tilde m} \partial_\tau \tilde x^{\tilde m}
-\frac {e}{2} \left(\tilde P_{\tilde m} \tilde g^{\tilde m\tilde n} \tilde
P_{\tilde n} + m^2\right)\right]\,
\label{particleaction}
\end{equation}
and also in a clearly reparametrization invariant form:
\begin{equation}
S = \int d\tau \left[ \sqrt g g^{\tau \tau}
\partial_\tau \tilde x^{\tilde m} \partial_\tau \tilde x^{\tilde n}
\tilde g_{\tilde m\tilde n} -\frac {\sqrt g}{2} m^2 \right]\, .
\label{particleaction2}
\end{equation}
One can gauge-fix the reparametrization symmetry but to the best of our
understanding, there is no clear way to make the action quadratic using any
of these forms of the classical action. The only exception is the
massless case. Note that the actions (\ref{particleaction}) and
(\ref{particleaction2})
consistently contain the limit
 $m\rightarrow 0$, as opposed to
(\ref{particlesimple}).
For $m=0$,  by going to the conformally flat metric of $AdS_5$,
the factor $\tilde g_{\tilde m \tilde n}$ in (\ref{particleaction2}) can be
completely absorbed  into the worldline metric, and the action becomes
quadratic upon gauge fixing the reparametrization symmetry.
However, since the mass parameter prevents us from doing the same successfully
in the massive case,
our strategy is to first write the classical Lagrangian of a massive particle
in $AdS_5$
using different field variables.

A good candidate for a set of coordinates are twistors.
These variables  are  well defined in the context of the
4-dimensional space \cite{Penrose,Ferber} and realize the conformal
symmetry $SO(2,4)\sim SU(2,2)$ linearly. Since this is also the
isometry group of $AdS_5$ (with 4d Minkowski space as its boundary)
it seemed natural to extend the twistor construction to include the
radial coordinate $\rho$. This was achieved in \cite{CRZ}.
For this purpose two twistors
\be
{\cal Z}^I=\pmatrix{\lambda_\a^I \cr \bar \mu^{\da I}}\, .
\ee
with $I=1,2$ needed to be introduced instead of one.
The space-time momenta $P$ are then replaced by the bilinears of the upper
components of the twistors, $\lambda$
\begin{eqnarray}
P_{\a\da} &=& 2 \lambda_\a^I \bar\lambda_\da^I\,\nonumber\\
P_\rho &=& -\frac i{2\rho^2} \left(\varepsilon^{\a\b} \epsilon^{IJ}
\lambda_\a^I \lambda_\b^J - c.c.\right)\, .
\label{momenta}\end{eqnarray}
The 5d space-time coordinates $x^m,\rho$ are encoded in the twistor
construction through the relation between the upper and lower
twistor components:
\be
\bar \mu^{\da I}=-i x^{\da \a}\lambda_\a^I+\frac{\eps^{IJ}}{\rho}
\bar \lambda^{\da J}\,
 \label{RelationAdS}
\ee
with $\e^{12}=1$.
(\ref{RelationAdS}) automatically implies
3 real constraints the twistors have to satisfy:
\begin{equation}
\bar {\cal Z}^I \hat \sigma_a^{IJ} {\cal Z}^J = 0\,,
\label{twistorconstraints}
\end{equation}
where $\sigma_i^{IJ}$ are the Pauli matrices, and where
\be
\bar{\cal Z}^I=({\cal Z}^I)^\dagger H, \qquad {\rm with} \qquad H=\pmatrix{0 &
1\cr 1& 0}\,
\label{conjugateTwistor}
\ee
is the conjugate twistor.

This new twistor construction for $AdS_5$ was then applied \cite{CRZ}
to the dynamics of a massive particle propagating on this space.
The action (\ref{particleaction}) in standard two-component
notation (see \cite{CRZ} for detailed conventions) is given by
\begin{equation}
S = -\int d\tau\,\left[ \frac12 P_{\a\da} \dot x^{\da\a} - P_\rho \dot \rho
- \frac{e}{2R^2}
\left( \frac{1}{2\rho^2} P_{\a\da} P^{\da\a} - \rho^2 P_\rho^2 -
m^2 R^2\right)\right]\, .
\label{2comp}
\end{equation}
Using relations (\ref{momenta}) and (\ref{RelationAdS}) one can rewrite
the action (\ref{2comp}) as follows
\begin{equation}
S = - i \int d\tau\, \bar {\cal Z}^I \partial_\tau{\cal Z}^I -
\frac{e}{2R^2} \left [ {1\over 4} (\bar {\cal Z}^I \delta^{IJ} {\cal Z}^J)^2
- ( m R )^2\right ] \,,
\label{twistoraction}
\end{equation}
where the twistor pair is subject to constraints (\ref{twistorconstraints}).
The reparametrization constraint, the variation of the action
over the worldline metric,
implies a fourth constraint
\begin{equation}
{1\over 2} |\bar {\cal Z}^I \delta^{IJ} {\cal Z}^J| = m R\, .
\label{Constraint4}
\end{equation}
All four constraints are quadratic in the twistors, and the classical action
can be brought into the form \cite{CRZ}
\begin{equation}
S_{cl} = - i \int d\tau\, \left(\bar {\cal Z}^I \partial_\tau{\cal Z}^I
-i u^a \phi_a\right) \,,
\label{twistoractionC}
\end{equation}
where $u^a (\tau)$ are 4 Lagrange multipliers to the 4 constraints
\begin{equation}
\phi_a = \bar {\cal Z}^I t_a^{IJ} {\cal Z}^J - 2 \delta_a{}^0 s R\, , \qquad
t_a^{IJ} = \left\{\delta^{IJ}, (\hat \sigma_i)^{IJ}\right\} \,
\end{equation}
and $s=\pm m$.
Note that there is an ambiguity in imposing the constraint
$\phi^0$. The original particle action induces (\ref{Constraint4}),
hence $\phi^0$ can be stated with $\mp 2mR$. Therefore, we have the
situation that two different twistor actions are equivalent to one particle
action. Fortunately, we will also see that both twistor theories are related
by a global symmetry, and therefore equivalent.

This classical twistor Lagrangian (\ref{twistoractionC}) is equivalent to the
original classical particle action in $AdS_5$ space since both theories
result in the same classical equations of motion.
The twistor Lagrangian has  an $U(1)\times SU(2)$ gauge symmetry
which acts on the fields as
\begin{eqnarray}
\delta {\cal Z}^I &=& i \xi^a(\tau)  t_a^{IJ} {\cal Z}^J\,,\nonumber\\
\delta u^0 &=& \partial_\tau \xi^0(\tau) \,,\nonumber\\
\delta u^i &=& \partial_\tau \xi^i(\tau)
               + 2 \varepsilon^{ijk} u^j \xi^k(\tau)\,,
\label{sym}
\end{eqnarray}
with local parameters $\xi^a(\tau)$. Also, the action has a global $SU(2,2)$
symmetry since any bilinear combination of twistors (or their space-time
derivatives) of the form
\begin{equation}
\bar {\cal Z}^I (\tau_1)
{\cal Z}^J (\tau_2) = \mu^{\alpha I} (\tau_1) \lambda_\alpha^J (\tau_2)  +
\bar\lambda_\da^I (\tau_1)  \bar\mu^{\da J} (\tau_2) \
\label{AdSinvariant}
\end{equation}
is invariant under the global $SU(2,2)$
\begin{equation}
\delta {\cal Z}^I = -g  {\cal Z}^I = - (\epsilon^\Lambda T_\Lambda)
{\cal Z}^I,
\end{equation}
where the transformations are those of the fundamental representation.

The 4 gauge symmetries of the action are consequences of the fact that
the twistor pair is subject to the four real first class constraints.
The canonical Poisson brackets $[q,p]_{P.B.} = 1$  following  from the
twistor action \eqn{twistoraction} are
\begin{equation}
[{\cal Z}^I_A, \bar {\cal Z}^J_B]_{P.B.} = i \delta^{IJ} \delta_{AB}\,,\quad
[{\cal Z}^I_A, {\cal Z}^J_B ]_{P.B.}
= [\bar {\cal Z}^I_A, \bar {\cal Z}^J_B ]_{P.B.} = 0\,,
\label{PB}
\end{equation}
where $A,B$ label the four twistor components $\lambda_\alpha, \bar
\mu^{\da}$.
With (\ref{PB}) it is easy to verify that the constraints form an
$U(1)\times SU(2)$ algebra
\begin{equation}
[\phi_0, \phi_i]_{P.B.} =  0\,,\qquad [\phi_i, \phi_j]_{P.B.} = -2
\varepsilon_{ijk} \phi_k\,.
\label{algebraconstr}
\end{equation}
At this point it is instructive to compare the degrees of freedom
of (\ref{particleaction}) and (\ref{twistoractionC}).
The original space-time action had 5 degrees of freedom
$(\tilde x^{\tilde m}, \tilde P_{\tilde m})$ and one gauge symmetry
(or one first class constraint), leading to
one pair of Faddeev-Popov ghosts. The net number of physical
degrees of freedom is $5-1=4$. The twistor action has 8 degrees of
freedom ${\cal Z}^I , \bar {\cal Z}^I$ and 4 gauge symmetries
(or 4 first class constraints). This leads to four pairs of Faddeev-Popov
ghosts, and the net number of physical degrees of freedom is $8-4=4$.
Thus the total number of physical degrees of freedom in both action
coincides which provides an argument that they are equivalent
quantum-mechanically.

The quantization can be performed both in the Lagrangian form in the field
space, as well as in  canonical form. The first one allows to find a gauge
in which the gauge-fixed action is quadratic, showing that the theory is free.
The BRST symmetry of this action proves that the path integral is
gauge-independent.

The operator form of the quantization in the canonical space has the advantage
that one can construct the BRST operator so that the Hamiltonian is given by
the commutator of the BRST operator with the gauge fermion $\Psi$,
\begin{equation}
H= [Q_{BRST}, \Psi] \quad \Rightarrow \quad [H,Q_{BRST}]=0
\end{equation}
and therefore the Hamiltonian commutes with the BRST operator
by construction. This leads to a simple definition of the physical
states in terms of the oscillators of the quantized theory.
It is particularly important in our case to find both, the generators of the
global $SU(2,2)$ symmetry, as well as the BRST operator which is the global
quantum counterpart of the $U(1)\times SU(2)$ local gauge symmetry in terms
of the free oscillators of our theory.  Since the action
(\ref{twistoractionC})
realizes the symmetry $SU(2,2)$ linearly we would expect that the states
of the theory fall into $SU(2,2)$ representations.
In \cite{GUNSAC,v} these representations (more precisely, those of
$SU(2,2|4)$) were constructed by an oscillator method. We will later see
that a subset of their representations built the Fock space of
our theory.

In order for their construction to be valid in our quantum theory
one has to verify that all $SU(2,2)$ generators commute with the
BRST operator $Q_{BRST}$. We will find that this is indeed the case.

As our theory has constraints, we will use Dirac's methods of
quantization of constrained systems \cite{Dirac}. These methods in
the modern form in application to the path integral were developed by
Fradkin and his collaborators \cite{BFVFT} and by Becchi, Rouet, Stora and
Tyutin \cite{BRST}.

To proceed with the Lagrangian BRST quantization we add  gauge
fixing conditions through gauge functions $\chi^a$. For simplicity
we take the $\chi^a$  to be functions only of the classical fields
(Lagrange multipliers and twistors), but not of their derivatives.
The gauge fixed action is
\begin{equation}
{\cal L}_{g.f.} (\Phi_{cl} , b_a, c^a, \pi_a) =
{\cal L}_{cl} (\Phi_{cl}) + Q \Psi \,.
\qquad \Phi_{cl}= \{u, {\cal Z}, \bar {\cal Z}\}
\label{g.f.}
\end{equation}
where $\Psi$ is the gauge fermion of the form
\begin{equation}
\Psi = b_a \chi^a (\Phi_{cl})\,
\end{equation}
and the $b_a$ are anti-ghosts. The BRST action on the fields of
the classical action is defined as follows. On the fields of the classical
action it is given by eqs. (\ref{sym}) with the parameter of the gauge
transformation $\xi^a(\tau)$ replaced by a product of the ghost field and a
global anticommuting parameter $\Lambda$.
In addition we have to specify the standard BRST transformation on
ghosts, anti-ghosts and multipliers to the gauge fixing condition $\pi_a$
\begin{eqnarray}
\delta_{BRST} \Phi_{cl} &=& \delta_{cl}
\Phi_{cl}(\xi^a(\tau)\rightarrow c^a(\tau )  \Lambda)  ,\nonumber\\
\delta_{BRST} b_a&=& \pi_a \Lambda \nn \\
\delta_{BRST} c^a&=& -{1\over 2} f_{bc}^a c^b c^c \Lambda \\
\delta_{BRST} \pi_a&=&0 \nn
\label{BRST}\end{eqnarray}
Using the notation $\delta_{BRST} X = QX \Lambda$ we can easily
verify that $Q^2=0$. Therefore, the action above is BRST invariant
since the classical action is invariant under BRST  transformations
and the gauge fixing term is invariant due to the nilpotency of $Q$.
\begin{equation}
Q {\cal L}_{g.f.} = Q {\cal L}_{cl} + Q^2 \Psi =0\,.
\label{g.f.1}
\end{equation}
The BRST invariant action  in this class of gauges can be rewritten as
\begin{equation}
{\cal L}_{g.f.} = {\cal L}_{cl} (\Phi_{cl}) + \pi_a \chi^a ( \Phi_{cl})
+ b_a { \partial  \chi^a ( \Phi_{cl}) \over \Phi_{cl}} Q \Phi_{cl}\,.
\label{g.f.2}
\end{equation}
The path integral is independent on the choice of the gauge-fixing function
$ \chi^a ( \Phi_{cl})$. With the simple choice
\begin{equation}
\chi^0= u^0-1 \qquad \chi^i= u^i \,
\end{equation}
the action becomes
\begin{equation}
{\cal L}_{g.f.} = - i  \bar {\cal Z}^I \partial_\tau{\cal Z}^I -
u^0 (\bar {\cal Z}^I \delta^{IJ} {\cal Z}^J -s R ) + \tilde \pi_i u^i
+ \pi_0 (u^0 - 1) + b_a \partial_\tau c^a\,.
\end{equation}
The cubic terms in the classical action and in the ghost action proportional
to $u^i$ have been absorbed by a redefinition of $\pi_i$
into $\tilde \pi_i$. The dependence of the classical action on $u^0(\tau)$
is not absorbed into the redefinition of $\pi_0$.
Since this is a reparametrization related symmetry,
and $u^0$ is related to $e$, it is not attractive to gauge away the
worldline metric, but rather to choose a gauge which sets $u^0(\tau)$
to a constant. The last step is simply to integrate over $\tilde \pi_i$
and $\pi_0$, or equivalently to solve the equations for these variables.
The final form of the gauge-fixed action in this gauge is quadratic in all
fields
\begin{equation}
S_{g.f.} = - i\int d\tau\,   \bar {\cal Z}^I \partial_\tau{\cal Z}^I -i
(\bar {\cal Z}^I \delta^{IJ} {\cal Z}^J - 2 s R ) + i
b_a \partial_\tau c^a\,.
\label{quadratic}
\end{equation}
This shows that the resulting theory is free and we may use
oscillators to construct the spectrum. The BRST symmetry is inherited from
the general case described above since we
have just taken a particular gauge and excluded the auxiliary fields.
We may also perform the quantization in  the elegant and powerful
formalism of Batalin and Vilkovisky \cite{Batalin}. This can be found in
the appendix.
From (\ref{quadratic}) one again derives the Poisson brackets
\begin{equation}
[{\cal Z}^I_A, \bar {\cal Z}^J_B]_{P.B.} = i \delta_{AB}
\delta^{IJ}\,.
\end{equation}
One might be tempted to identify ${\cal Z}^I$ ($\bar {\cal Z}^I$)
as creation (annihilation) operators, but we will wait with the
construction of states (i.e. the definition of a vacuum and creation and
annihilation operators) until later. In fact, we will find that there is
a more suitable choice.

Let us first construct the $SU(2,2)$ generators explicitly in terms of the
quantized twistors. We start from the classical action
\eqn{twistoraction} and will obtain the SU(2,2) generators by a
Noether procedure.  The twistors transform in the fundamental
representation of $SU(2,2)$ with $\delta {\cal Z}^I = - (\epsilon^\Lambda
T_\Lambda) {\cal Z}^I$.  The fundamental of $SU(2,2)$ is the (chiral)
spinor representation of $SO(4,2)$ and the generators are the $SO(2,4)$
gamma-matrices,
$
\hat M_{\hat m\hat n} = \frac 14 \hat \gamma_{\hat m\hat n}
$. We choose
$
\hat\gamma_{mn} = \gamma_{mn}\,,\ \hat \gamma_{mS} = \gamma_m\gamma_5
\,,\ \hat \gamma_{mT} = -\gamma_m\,,\ \gamma_{TS} = \gamma_5\,,
$
where the $\gamma_m,\gamma_5$ are 5-dimensional gamma-matrices. The
translation between the spinor representation of $SO(2,4)$ and the
fundamental of $SU(2,2)$ goes through the unitary similarity matrix which
relates the hermitian conjugate $\gamma$-matrices to the original ones
\begin{equation}
(\hat \gamma_{\hat m\hat n})^\dagger = -\hbox{\mybb A} \hat \gamma_{\hat m\hat
n} \hbox{\mybb A}^{-1}\,,\qquad \hbox{\mybb A}^2 = 1\,.
\end{equation}
The matrix $\hbox{\mybb A}$ is identified with the $SU(2,2)$ metric $H$ and is
given by
\begin{equation}
\hbox{\mybb A} = -i \gamma_0\,.
\end{equation}
It can be checked that in the twistor basis for the gamma-matrices
\cite{CRZ} it gives the metric as in \eqn{conjugateTwistor}.
Therefore the variation of the twistor, which is in the first place an
$SO(2,4)$-spinor is
\begin{equation}
\delta {\cal Z}^I = -\frac 14 \hat \Lambda^{\hat m \hat n} \hat \gamma_{\hat
m\hat n} {\cal Z}^I\,.
\end{equation}
In general, the variation of the action is
\begin{equation}
\delta S = \int d\tau \,(\partial_\tau \epsilon^\Lambda) J_\Lambda
\end{equation}
for local parameters $\epsilon^\Lambda$. The $J_\Lambda$ are the conserved
currents (in this case, as we deal with a quantum mechanical system,
they are right away the charges). Applying this to the case at hand
\begin{equation}
\delta S \ =\
\frac i4 \int d\tau\, (\partial_\tau \hat \Lambda^{\hat m\hat n})
\bar {\cal Z}^I \hat\gamma_{\hat m\hat n} {\cal Z}^I\,.
\end{equation}
We derive the $SU(2,2)$ charges in terms of twistors
\begin{equation}
\hat M_{\hat m\hat n} = \frac i4 \bar {\cal Z}^I \hat\gamma_{\hat m\hat n}
{\cal Z}^I\,,
\end{equation}
These are the 15 hermitian generators of $SU(2,2)\sim SO(2,4)$. The
Noether procedure guarantees that they satisfy the algebra of Poisson
brackets. To elevate this to the quantum theory,
we promote the Poisson brackets defined above to the commutators
\begin{equation}
[{\cal Z}^I_A, \bar {\cal Z}^J_B]_{P.B.} = i \delta_{AB} \delta^{IJ} \qquad
\Rightarrow \qquad
[\bar {\cal Z}^I_A, {\cal Z}^J_B] = \delta_{AB} \delta^{IJ}\, ,
\label{quant}\end{equation}
In terms of components the commutation relations are
\begin{equation}
[\mu^{\a I}, \lambda_\b^J] = \delta^\a{}_\b \delta^{IJ}\,,\qquad
[\bar \lambda_\da^I , \bar \mu^{\db J}] = \delta_\da{}^\db \delta^{IJ}\,.
\label{quant1}
\end{equation}
We will also have to deal with normal ordering ambiguities, but first we would
like to give the explicit correspondence between this twistor construction
and the oscillator construction of \cite{v}.

To construct the Fock space we have to identify the annihilation and
creation operators. We have found in (\ref{quant}) and (\ref{quant1})
that one can view $\lambda$ and $\bar \mu$ as creation operators, or any
linear combination of the two (the same holds, of course, for the annihilation
operators).
The commutation relations of our dynamical system are closely
related to those of the oscillator method used in \cite{v} to construct the
UIR's of $SU(2,2)$. Therefore, we would like to make use of the results
obtained there.

The oscillator construction is built upon a
set of creation (annihilation) operators $a^r,b^r$ $(a_r,b_r)$ satisfying
$a_r=(a^r)^\dagger, b_r=(b^r)^\dagger$. As the twistors as such do not
satisfy this condition a change of basis is required. Thus, we obtain the
following picture:
$$
\begin{array}{ccccc}
AdS\mbox{-spacetime}  &\Longrightarrow&  \mbox{twistors} &  \Longrightarrow
& \mbox{oscillator construction}\\[12pt]
(x^m, \rho) &\Longrightarrow& {\cal Z}^I=\pmatrix{\lambda_\a^I \cr \bar
\mu^{\da I}}
&\Longrightarrow &\Psi^I = \ba{c} a^{rI} \\ - b_s^I\ea\,.
\end{array}
$$
To explain the second part of this connection we make the change of basis
explicit. The change of  basis between $\gamma$-matrices in the oscillator
basis as in \cite{v}, denoted by $\gamma'$, and the twistor basis, denoted
by $\gamma$, is given by the following unitary similarity transformation
\begin{equation}
\hat \gamma'_{\hat m\hat n} = \hbox{\mybb S} \hat \gamma_{\hat m\hat n}
\hbox{\mybb S}^\dagger\,,
\end{equation}
where
\begin{equation}
\hbox{\mybb S} = \frac1{\sqrt 2} \ba{cc} \unity_2 & \unity_2 \\ -\unity_2 &
\unity_2\ea\,.
\end{equation}
This induces the following rotation on the twistors
\begin{equation}
\Psi^I = \hbox{\mybb S} {\cal Z}^I \,, \qquad \bar \Psi^I = \bar {\cal Z} ^I
\hbox{\mybb S}^\dagger\,.
\label{twisttoosc}
\end{equation}
Now we identify the oscillators in the spinors $\Psi$ \cite{v} as
\begin{equation}
\Psi^I = \ba{c} a^{rI} \\ - b_r^I\ea\,,\qquad
\bar \Psi^I \equiv -i \Psi^\dagger \gamma'_0 = \ba{cc}
a_r^I & b^{rI}\ea\,.
\end{equation}
The commutation relations for th $\Psi^I$ follow from the ones of
${\cal Z}^I$
\begin{equation}
[\bar \Psi^I_A, \Psi^J_B] = \hbox{\mybb S}_A{}^C [\bar {\cal Z}^I_C, {\cal
Z}^J_D]\, \hbox{\mybb S}^\dagger{}^D{}_B
= \delta_{AB} \delta^{IJ}\,,
\end{equation}
which in more detail read
\begin{equation}
[a_r^I, a^{s J}] = \delta_r{}^s \delta^{IJ}\,,\qquad
[b_r^I, b^{s J}] = \delta_r{}^s \delta^{IJ}\,,\qquad
\end{equation}
as desired. In this basis the $SU(2,2)$ metric $H=\hbox{\mybb A}$ is given
by $\mbox{diag}(1,1,-1,-1)$.
The global $SU(2,2)$ generators are
\begin{equation}
\hat M_{\hat m\hat n} =  \frac i4 \bar\Psi^I \hat \gamma'_{\hat m\hat n}
\Psi^I
\end{equation}
in the oscillator basis, and the gauge-fixed action \eqn{quadratic} reads
\begin{equation}
{S}_{g.f.} = -i\int d\tau\,  \bar \Psi^I \partial_\tau \Psi^I -i
\left(\bar \Psi^I \delta^{IJ} \Psi^J - 2sR\right) +i b_a \partial_\tau c^a\,.
\label{quadratic2}
\end{equation}

Now, that we have related the quantized twistors ${\cal Z}^I$ to the creation
$(a^r,b^r)$ and annihilation $(a_r,b_r)$
operators of \cite{v} we have proven that the oscillator construction of
UIR of $SU(2,2)$ in \cite{v} follows from the quantization of our action.
Before we revisit this construction let us first reexpress the $SU(2,2)$
generators explicitly in the $a,b$ basis. Since we want to define them as
quantum operators we have to tackle the problem of normal ordering.
In the following the index $\hat m$ takes values $\{0,l,5,6\}$,
where $l=1,2,3$.  We find that the following generators build the $SU(2,2)$
algebra on the quantum level:
\begin{eqnarray}
:\!\hat M_{lm}\!: &=& -\frac14 \varepsilon_{lmn} \left[a^{r I} a_s^I (\hat
\sigma_m)^s{}_r - b^{rI} b_s^I (\hat \sigma_m)_r{}^s \right]\,,\nn\\
:\!\hat M_{l5}\!: &=& -\frac14 \left[ a^{rI} a_s^I (\hat \sigma_l)^s{}_r
+ b^{rI} b_s^I (\hat \sigma_l)_r{}^s\right]\,,\nn\\
:\!\hat M_{06}\!: &=& \frac14 \left[a^{rI} a_r^I + b^{r I} b_r^I + 4\right]\,,
\nn\\
:\!\hat M_{0l}\!: &=& \frac i4 \left[a^{rI} b^{sI} (\hat\sigma_l)_{sr} - a_r^I
b_s^I
(\hat\sigma_l)^{rs}\right]\,,\nn\\
:\!\hat M_{6l}\!: &=& \frac14 \left[a^{rI} b^{sI} (\hat\sigma_l)_{sr} + a_r^I
b_s^I
(\hat\sigma_l)^{rs}\right]\,,\nn\\
:\!\hat M_{05}\!: &=& \frac14 \left[a^{rI} b^{sI} \delta_{rs} + a_r^I b_s^I
\delta^{rs}\right]\,,\nn\\
:\!\hat M_{65}\!: &=& \frac i4 \left[a^{rI} b^{sI} \delta_{rs} - a_r^I b_s^I
\delta^{rs}\right]\,.
\label{quantumgenSU22}
\end{eqnarray}
The generators $:\!\hat M_{l5}+\frac{1}{2}\eps_{lmn}
M_{mn}\!:$, $:\!\hat M_{l5}-\frac{1}{2}\eps_{lmn} \hat M_{mn}\!:
$, $:\!\hat M_{06}\!:$ generate the maximal compact subgroup
$SU(2)_L\times SU(2)_R \times U(1)_E$, which provides the quantum numbers
of the states. The operator
$:\!\hat M_{06}\!:$ corresponds to the $AdS$ energy and generates translations
along the global
timelike vector field of the $AdS_5$ space.  In the conformal
interpretation it is the conformal Hamiltonian, $\frac 12 (P_0 + R^2 K_0)$.
This generator is the only global $SU(2,2)$ generator which suffers from a
normal ordering ambiguity, because of the tracelessness of the Pauli
matrices.  This ambiguity can be resolved by demanding
that the $SU(2,2)$ is a global symmetry of the quantum theory and since we
are dealing with a simple algebra (therefore every operator appears on the
r.h.s.~of the algebra) the quantum operators are given necessarily by
\eqn{quantumgenSU22}.  These satisfy the algebra of commutation relations
\begin{equation}
[\, :\! \hat M_{\hat m\hat n}\!: , :\! \hat M_{\hat p\hat q}\!:\,]  =
-i (\hat \eta_{\hat m[\hat p}
:\!\hat M_{\hat q]\hat n}\!: - \hat \eta_{\hat n[\hat p}
:\!\hat M_{\hat q]\hat m}\!:)\,.
\end{equation}
For convenience, we introduce the number operators
\begin{equation}
N_a = a^{r I} a_r^I\,,\qquad
N_b = b^{r I} b_r^I\,.
\label{counting}
\end{equation}
The $AdS$ energy
\begin{equation}
E = \frac 2R :\! \hat M_{06}\!:\,  = \frac 1{2R} (N_a + N_b +4)
\label{energy}
\end{equation}
is quantized in units of $R^{-1}$.

The Fock space of (zero ghost number) states is constructed  as in \cite{v}
with just two generations of oscillators. The vacuum is defined by
\be
a_r|0\rangle=b_r|0\rangle=0\, .
\ee
The lowest weight type representations (lwtr) are then constructed
by selecting a lowest weight vector (lwv) $|\Omega\rangle$ annihilated by all
$SU(2,2)$ generators of the form
\be
L^-_{rs}\equiv (a_r)^I (b_s)^I\,.
\ee
The lwv transforms irreducibly under the maximal compact subgroup
$S(U(2)\times U(2))$,  $ L^0\equiv (a^r)^I (a_s)^I \oplus (b^r)^I (b_s)^I$,
and the representations are then constructed by the repeated action of
\be
L^+{}^{rs} \equiv (a^r)^I  (b^s)^I
\ee
on $|\Omega\rangle$, i.e. the representations are of the form
\be
|\Omega\rangle,\quad L^+|\Omega\rangle, \quad L^+ L^+|\Omega\rangle,\quad L^+
L^+L^+|\Omega\rangle,..
\ee

Finally, let us proceed with the Hamiltonian quantization.
The physical states of the theory can now be constructed using the
BRST operator. As usual, for the first class constraints $[\phi_a , \phi_b] =
f_{ab}^c \phi_c$  one defines $Q_{BRST}= c^a \phi_a - {1\over 2} f_{ab}^c c^a
c^b
b_c$, which in our case becomes
\begin{equation}
Q_{BRST} = c^a \phi_a +  i \varepsilon_{ijk} c^i c^j  b_k\, .
\label{Qbrst}
\end{equation}
It is easily verified that one has indeed $Q_{BRST}^2=0$. There are no normal
ordering issues, since the only expression which could lead to such
ambiguities,
$\phi_0$, drops out out of the calculation of $Q_{BRST}^2$.

The physical states $|\Omega\rangle$ of the theory must also be constructed in
a way that respects the global symmetry of our action.
Fortunately, by construction, our BRST operator commutes with the generators
of the global $SU(2,2)$, because the constraints in $Q_{BRST}$ contain the
twistors only in the $SU(2,2)$ invariant bilinear form of
(\ref{AdSinvariant}).
Therefore, the construction of representations above is consistent with
the BRST condition on physical states $|\Omega\rangle$ given by
\begin{equation}
Q_{BRST} |\Omega \rangle = 0 \qquad |\Omega \rangle \neq Q_{BRST}
|\tilde \Omega \rangle
\label{condition}
\end{equation}
As usual, we will focus here on the states with vanishing ghost number
\begin{equation}
Q_{BRST} |\Omega \rangle_{N_{gh}=0} = c^a \phi_a
|\Omega \rangle_{N_{gh}=0} =0,
\end{equation}
which are by construction annihilated by the (normal ordered) BRST constraints
\begin{eqnarray}
:\!\phi_0\!: &=& :\!\bar {\cal Z}^I {\cal Z}^I -  2 s R\!:
\ =\ :\! \bar \Psi^I \Psi^I - 2 sR\!:
\ =\   (N_a - N_b + 4 - 2(s+s_0)R )\,,\nn\\
:\!\phi_i\!:&=& :\!\bar {\cal Z}^I (\hat \sigma_i)^{IJ} {\cal Z}^J\!: \ =\
:\!\bar \Psi^I
(\hat\sigma_i)^{IJ} \Psi^J\!: \ =\
a^{rJ} a_r^I (\hat\sigma_i)^{IJ} - b^{rI} b_r^J (\hat \sigma_i)^{IJ}
\end{eqnarray}
In the $SU(2)$ constraints $:\!\phi_i\!:$ there is no normal ordering
constant, because the Pauli matrices are traceless.
The nonvanishing commutators of above constraints are
\be
[:\!\phi_i\!:,:\!\phi_j\!:]= - 2 i\varepsilon_{ijk}:\!\phi_k\!:\,.
\ee
However, $:\!\phi_0\!:$ has a normal ordering ambiguity expressed in
the new parameter $s_0$. This ambiguity can neither be resolved by
requiring nilpotency of the BRST operator, nor closure of the algebra.
Hence, the theory seems to be consistent for all choices of $s_0$.

However the classical twistor action \eqn{twistoraction} and the
gauge-fixed actions \eqn{quadratic} and \eqn{quadratic2} are
\lq\lq{\sf CPT} invariant''\footnote{We mean {\sf CPT} invariance in the
4-dimensional sense as in \cite{v}.} for the transformations
\begin{equation}
\Psi^I \stackrel{{\sf CPT}}{\longrightarrow} - \gamma'_5 (\Psi^I)^*\,,\qquad
\Longrightarrow\qquad
a \stackrel{{\sf CPT}}{\longrightarrow} b\,,\quad
b \stackrel{{\sf CPT}}{\longrightarrow} a\,,
\end{equation}
since $\Psi^I$ is a Dirac spinor, and
\begin{equation}
\tau \stackrel{{\sf CPT}}{\longrightarrow} -\tau\,,\quad
c^a \stackrel{{\sf CPT}}{\longrightarrow} c^a\,, \quad
b_a \stackrel{{\sf CPT}}{\longrightarrow} b_a\quad \mbox{and} \quad
s \stackrel{{\sf CPT}}{\longrightarrow} -s\,.
\end{equation}
We see that under a \lq\lq {\sf CPT}'' transformation the parameter s
necessarily changes sign. If we want to extend this symmetry to a quantum
symmetry and construct a \lq\lq {\sf CPT} invariant'' spectrum we need that
the constraints transform into themselves under this {\sf CPT}. This
leads us to $s_0=2$ as then under {\sf CPT} the constraint
$:\!\phi_0\!:\ \stackrel{{\sf CPT}}{\longrightarrow}\ - :\!\phi_0\!:$.
This constraint then gives that on physical states
\begin{equation}
Z \equiv sR= {1\over 2} (N_a - N_b)
\, .
\label{Z}
\end{equation}
Since this generator commutes with all $SU(2,2)$ generators all
states of a given representation carry the same eigenvalue of $Z$.
In the full supersymmetric case a $U(1)$ operator (central charge)
does appear in the r.h.s. of the superconformal algebra \cite{v},
which takes in the bosonic case precisely the form of $Z$. This suggests
that in fact the mass $mR$ is to be identified with the modulus of
the central charge in the superconformal theory. However, a full clarification
of this point requires a supersymmetric generalization of this work.

Now, let us discuss the representations of the theory which form the physical
states. We will not attempt a full classification in this
publication, rather have a first glance at the problem. We analyze
the representations of \cite{v} with two generations of oscillators
and check whether they are admitted by the BRST constraints (i.e.
$SU(2)$ invariance).

The first state allowed (with the choice $s_0=2$) is the vacuum state
\be
|\Omega\rangle=|0\rangle \, .
\ee
The corresponding  multiplet arises from the ``massless
(super)particle'' and gives rise (in the supersymmetric case)
to the supergravity multiplet of ${\cal N}=8$ supergravity.

Generically, $SU(2,2)$ representations with lwv
\be
|\Omega\rangle=a^{r_1I_1}a^{r_2I_2}...a^{r_nI_n}   |0\rangle \, ,
\ee
are not annihilated by the BRST operator, and therefore not
physical states of our theory.

To construct admissible representations is rather simple, as we only need to
use $SU(2)$ representation theory (where the $SU(2)$ is part of the
BRST symmetry and not be confused with any $SU(2,2)$ subgroup).
The oscillators transform in the fundamental
representation of $SU(2)$. For example
the Young tableaux of an $SU(2)$ invariant state is given by
\begin{equation}
\label{tab2}
\left|\lmarcshapirowbox ~, 1
\right\rangle \, ,
\end{equation}
\vskip.5cm
\noindent
representing the lwv
\be
|\Omega\rangle=\eps_{r_1r_2}\eps_{I_1I_1}...\eps_{r_{2j-1}r_{2j}}\eps_{I_{2j-1}I_{2j}}
a^{r_1I_1}a^{r_2I_2}a^{r_3I_3}a^{r_4I_4}...a^{r_{2j-1}I_{2j-1}}a^{r_{2j}I_{2j}}|0\rangle
\,.
\ee
The lwv of representations of this type have $AdS$ energies
\be
ER=2+{1\over 2}N_a= 2+j \, ,
\ee
where $N_a$ is a non-negative even integer,
and the entire multiplet has
\be
Z= {1\over 2}N_a= j .
\ee
These are the multiplets which
in the supersymmetric case were called in \cite{v} novel short multiplets.
The lwv of the ``{\sf CPT} conjugate'' multiplets \cite{v},
obtained by replacing $a$ with $b$ oscillators, have the same $AdS_5$
energy, but $Z=-j$.

Above we gave examples of $SU(2,2)$ UIR's which are physical states
of the theory, and of UIR's which are not. A complete study of the
admitted states remains to be done.

\bigskip

In conclusion, we have quantized the dynamics of
a massive particle propagating on $AdS_5$. We applied the BRST quantization
formalism to the action following from the twistor
parametrization of $AdS_5$. The action turned out to be free and we
specified the Fock space of the
theory. We found that the states fall into $SU(2,2)$ representations
with integral value of the ``central charge'' $Z$, corresponding to
the mass of the particle.
They correspond to the bosonic part of the
novel short supermultiplets \cite{v} of states on $AdS_5$.

Given the success of  twistor variables in the particle case we hope that the
procedure can be generalized and may help also in the
quantization of the string on $AdS_5$.

\vspace{1.5cm}

\noindent
{\bf Acknowledgements:} We had stimulating discussions with M. Gunaydin,
D. Minic, A. Peet, R.
Penrose, S. Shenker, A. Strominger,
A. Van Proeyen, Y. Zunger and C. Vafa.  The work of P.C. was supported by the
European Commission TMR program ERBFMRX-CT96-0045. R.K.~and J.R.~were
supported in part by NSF grant PHY-9870115.

\vspace{2cm}

\noindent
{\bf \large BV quantization of the particle on AdS$_5$ in a nutshell}

\vspace{0.5cm}
\noindent
In the BV method \cite{Batalin} we introduce for every field
(including the ghosts) an antifield with opposite statistics.
The space of fields and antifields are called the Fields.  We denote them
generically with $\Phi = \{\phi^A, \phi^*_A\}$.  The antifields are
assigned a ghostnumber $gh(\Phi)$ such that
\begin{equation}
gh(\phi^A) + gh(\phi^*_A) = -1\,.
\end{equation}
The various fields of our model are given in table \ref{tab:fields}.
\medskip

\begin{table}[h]
\begin{center}
\begin{tabular}{|c|c|c||c|c|c|}
\hline
\multicolumn{3}{|c||}{fields} & \multicolumn{3}{c|}{antifields}\\\hline
name& stat & $gh$ & name &stat & $gh$\\\hline\hline
${\cal Z}^I$ & + & 0 &$\bar {\cal Z}^*{}^I$ &-& -1\\
$\bar {\cal Z}^I$ & + & 0& ${\cal Z}^*{}^I$ &-& -1\\
$u^a$ & + & 0 & $u^*_a$ & - &-1\\
$c^a$ & - & 1 & $c^*_a$ & + &-2\\
\hline
\end{tabular}
\end{center}
\caption{The BV Fields of the model\label{tab:fields}}
\end{table}
\medskip
\noindent
One introduces the antibracket of two functions of the Fields $F,G$
\begin{equation}
(F,G) = \frac {F\stackrel{\leftarrow}{\partial}}{\partial \phi^A}\,
\frac {\stackrel{\rightarrow}{\partial}G}{\partial\phi^*_A} -
\frac {F\stackrel{\leftarrow}{\partial}}{\partial \phi^*_A}\,
\frac {\stackrel{\rightarrow}{\partial}G}{\partial\phi^A}\,.
\end{equation}
In the summation over $A$ also an integration over $\tau$ is understood.
The extended action for the model is
\begin{eqnarray}
S_{ext} &=& -i \bar {\cal Z}^I \partial_\tau {\cal Z}^I - u^a (\bar{\cal
Z}^I t_a^{IJ} {\cal Z}^J - 2 \delta_a{}^0 m R)\nonumber\\
&& + i \bar {\cal Z}^*{}^I t_a^{IJ} {\cal Z}^J c^a - i
\bar {\cal Z}^I t_a^{IJ} {\cal Z}^*{}^J c^a\nonumber\\
&& + u^*_a (\partial_\tau c^a + 2 \varepsilon^{ijk} u^j c^k \delta_i{}^a) +
c^*_i \varepsilon^{ijk} c^j c^k\,
\end{eqnarray}
and is obtained by requiring the classical master equation
\begin{equation}
(S_{ext}, S_{ext})=0\,.
\label{mastereq}
\end{equation}
The space of physical observables is defined through the antibracket
cohomology.  We have the nilpotent (through Jacobi-identities and
\eqn{mastereq}) operator
\begin{equation}
{\cal S} F = (F,S)\,.
\end{equation}
The physical observables are functions of the fields of ghostnumber 0 and
two functions describe the same physical observable if they are in the same
cohomology class.  We define canonicle transformations as transformations
which preserve the antibracket.  It is then clear that this does not affect
the physical observables.  It turns out that gauge-fixing can be performed
by a canonical transformation.  In our case we take the transformation
\begin{equation}
u^*_a = b_a\,,\qquad u^a = \delta_0{}^a - b^*{}^a\,,
\end{equation}
provides an appropriate gauge-fixing.
What it means is that we choose a different subset in the space of Fields of
what we call fields.  Indeed, we consider $u^*_a$ as a field and give it the
name $b_a$.  It has ghostnumber $-1$ hence its usual name antighost.
The gauge-fixed action gets the form
\begin{eqnarray}
S_{g.f.} &=& -i \bar {\cal Z}^I \partial_\tau {\cal Z}^I - (\bar{\cal
Z}^I {\cal Z}^I - 2 m R) + b_a \partial_\tau c^a\nonumber\\
&& + i \bar {\cal Z}^*{}^I t_a^{IJ} {\cal Z}^J c^a - i
\bar {\cal Z}^I t_a^{IJ} {\cal Z}^*{}^J c^a\nonumber\\
&& + 2 b^{*i} \varepsilon^{ijk} b^j c^k + c^*_i \varepsilon^{ijk} c^jc^k\,.
\end{eqnarray}
The antifield-independent part of this action coincides with action
\eqn{quadratic}.
The BRST operator $\Omega$ is identified through
\begin{equation}
\Omega f(\phi) = \left.(f(\phi), S)\right|_{\phi^*=0}\,.
\end{equation}

\end{document}